# Asymmetrical free diffusion with orientation-dependence of molecules in finite timescales

Nan Sheng[1,†], Yusong Tu[2,†], Pan Guo[1], Rongzheng Wan[1], Haiping Fang[1,*]

[1]*Division of Interfacial Water, Shanghai Institute of Applied Physics, Chinese Academy of Sciences, P.O. Box 800-204, Shanghai 201800, China*

[2] *Institute of Systems Biology, Shanghai University, Shanghai, 200444, China*

†*Equal contribution to this work*

*Corresponding author (email: fanghaiping@sinap.ac.cn)

**Using molecular dynamics simulations, we show that free diffusion of a nanoscale particle (molecule) with asymmetric structure critically depends on the orientation in a finite timescale of picoseconds to nanoseconds. In a timescale of ~100 ps, there are ~10% more possibilities for the particle moving along the initial orientation than moving opposite to the orientation; and the diffusion distances of the particle reach ~1 nm. We find that the key to this observation is the orientation-dependence of the damping force to the moving of the nanoscale particle and a finite time is required to regulate the particle orientation. This finding extends the work of Einstein to nano-world beyond random Brownian motion, thus will have a critical role in the understanding of the nanoscale world.**

Diffusion is one of the most critical transport phenomena that it enables mix or mass transport without requiring any extra-energy input. Free diffusion of particles from the molecular to the macroscopic scale suspended in liquids or gases has been extensively studied including the early work done by Einstein [1]. Tracking of particle diffusion is always considered as a result of the random walk of particles [2, 3], and thus the displacements of the particle are handled as isotropic. In recent years, it has become well-recognized that extensive biological, chemical and even physical processes including conformation changes usually occur in nano-, pico- and even femto-seconds [4-7], and very small spaces [8-15]. For example, the macromolecules in cells are typically separated by only 1-2 nm [5, 16-18] and dynamics of these macromolecules occur at the nanoscale. Until recently, conventional theories are still widely applied in studying the behavior in various systems with finite timescales and length scales, which may have led to misunderstanding. Recently, we have found the negative correlation between diffusion of ammonia molecule in water and the dipole orientation [19].

In this paper, using molecular dynamics simulations, we present the spontaneous asymmetric diffusion, which is orientation-dependent diffusion of asymmetric nanoparticles in finite timescales. Here, we use molecules including methanol, glycine to illustrate this idea, and we find the asymmetric part of the diffusion reaches ~10% of the total diffusion when the diffusion distances of the particle reach ~1 nm. Interestingly, the asymmetric part is saturated when the time is sufficiently large, which is negligibly small at the macroscopic-level. We find that the orientation-dependent diffusion results from the orientation-dependent damping force of the asymmetric nanoparticle together with a finite time required to regulate the particle orientation from the initial orientation. This finding extends the work of Einstein to the nanoscale beyond random Brownian motion since most of the particles have asymmetric structures. We expect that the orientation-dependent diffusion may have a critical role in the dynamics of bio-molecules in living cells and various nanoscale devices, such as chemical separation, sensing and drug delivery.

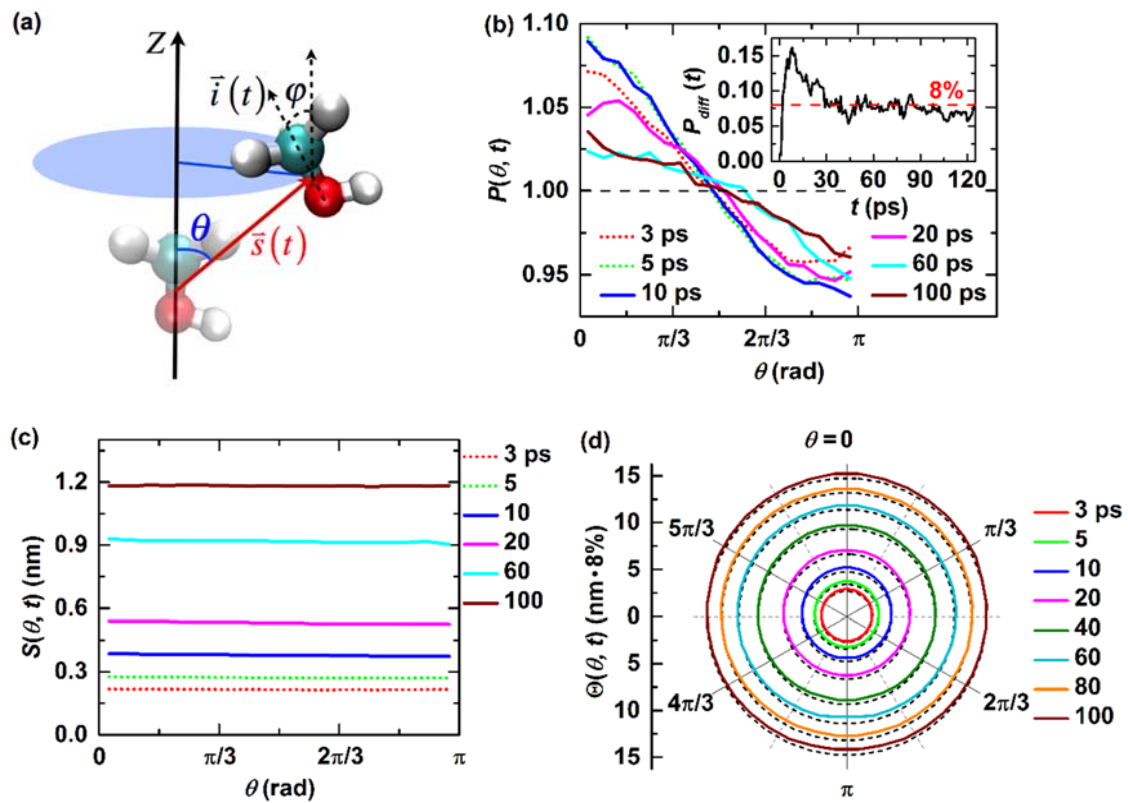

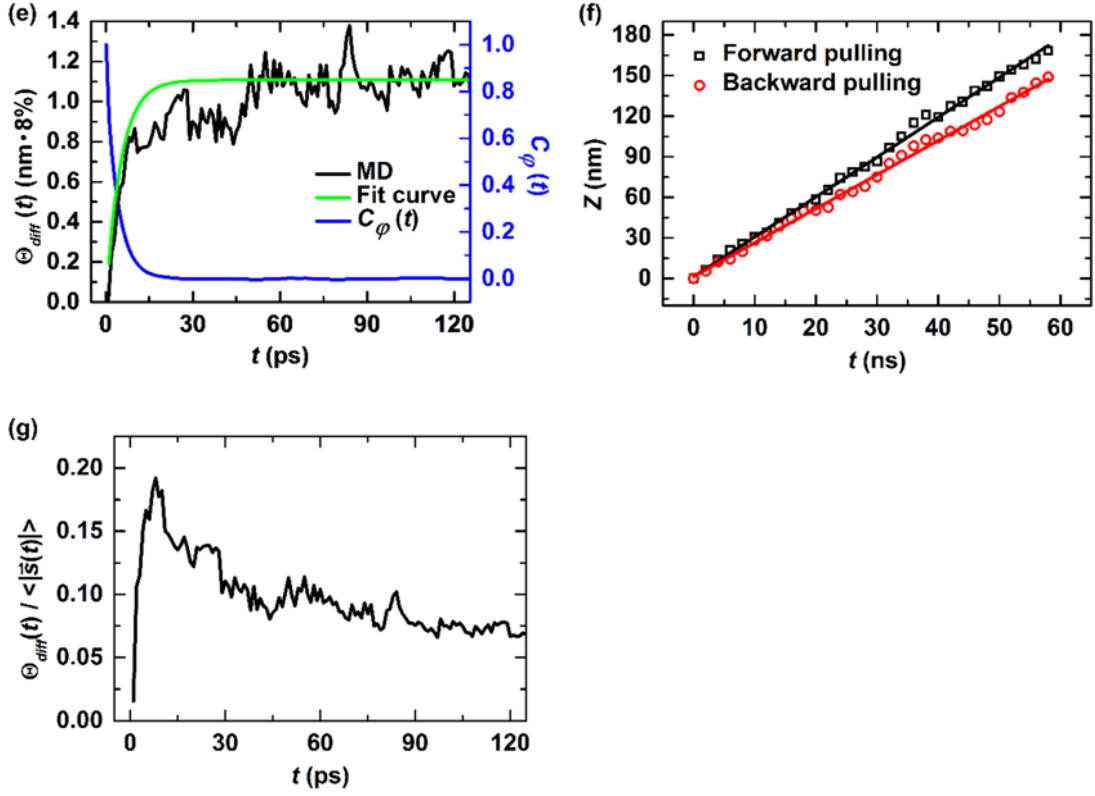

**Figure 1.** Asymmetric configuration of methanol and the orientation-dependent diffusion in finite timescales. **(a)** Configuration of methanol and the coordinate orientation. The oxygen, carbon and hydrogen atoms are shown by red, cyan, and white spheres, respectively. The coordinate is defined with the origin at the center of mass, and $z$ axis, along the initial orientation direction $\vec{i}(t=0)$ (pointing from the oxygen atom to the carbon atom) of the methanol molecule (shown by a transparent structure). $\vec{s}(t)$ is the displacement vector of the molecule at time $t$. $\theta$ is the position angle between $\vec{s}(t)$ and the $z$ axis. $\varphi$ is the orientation angle from $\vec{i}(t)$ to the $z$ axis. **(b)** Probability distribution $P(\theta,t)$ of the molecule displacements with respect to the position angle $\theta$. $P(\theta,t)$ is normalized by $\int_0^\pi P(\theta,t)d\theta/\pi = 1$. The inset is $P_{diff}(t) = P(0,t) - P(\pi,t)$. **(c)** Average displacement $S(\theta,t)$ of methanol molecule with respect to $\theta$ and time $t$. **(d)** Normalized accumulated displacement

$\Theta(\theta, t)$. $\Theta(\theta, t)$ = (*the average displacement at a position angle $\theta$ × the number of the samples with a position angle $\theta$ of the molecule*) × *a normalized factor* = $P(\theta, t) \cdot S(\theta, t)$, where $S(\theta, t)$ is the average displacement at $\theta$. $\Theta(\theta, t)$ represents the accumulated displacements of all the samples that the molecule displacements have a position angle $\theta$. $\Theta(\theta, t)$ is normalized to allow $\overline{\Theta}(t)$ equal to the average displacement $\langle |\vec{s}(t)| \rangle$, here $\langle \cdots \rangle$ denotes the average for all the samples, and $\overline{\Theta}(t) = \int_0^\pi \Theta(\theta, t) d\theta / \pi$. The left vertical axis marks the value of the displacement at $\theta = 0$. Different colors represents different time *t*. The dashed lines are circles with radii of $\overline{\Theta}(t)$ for guiding the eyes. In order to emphasize the contribution of the orientation-dependent part in the diffusion and considering that $P_{diff}(t)$ is ~8%, we use nm•8% as the unit of $\Theta(\theta, t)$ so that the length difference from $\Theta(\theta, t)$ to $\overline{\Theta}(t)$ is comparable to the diffusion displacement. (**e**) Difference of the maximal and minimal values of the normalized accumulated displacement, denoted by $\Theta_{\text{diff}}(t)$ (black), and the auto-correlation of molecule orientation angle $\varphi$, denoted by $C_\varphi(t)$ (blue). The green line is the exponential fit. (**f**) Z-position of the molecule pulled by a constant force along the +z and −z directions, respectively, indicating different damping coefficients for moving forward and backward. (**g**) Relative deviation of asymmetric diffusion from the average diffusion displacement $\Theta_{\text{diff}}(t) / \langle |\vec{s}(t)| \rangle$.

The simulation systems included one target molecule (methanol and glycine) solvated in 2178 water molecules. Parameters for target molecules were taken from the all-atom OPLS (optimized potentials for liquid simulations) force field [20] and TIP4P water model. We applied periodic boundary conditions in all directions, and the NVT ensemble at the temperature of 300 K by the Nosé-hoover thermostat [21] with a coupling coefficient of $\tau_P$ = 0.5 ps, using Gromacs 4.5 software [22]. A time step of 1.0 fs was used, and the neighbor list was updated every step within the cutoff radius of 1.6 nm. The Particle Mesh Ewald (PME) method [23] was used for long range electrostatics, whereas the switch function was used for both Coulomb for short range and van de Waals (vdW) interactions. Here, we performed 350 ns numerical simulation for each target molecule, and collected the frames for every 100 fs after the first 2 ns simulation. In the data analysis, each frame was set as the initial frame ($t = 0$) for one sample. The displacement in the next time interval $t$ is computed. Thus, for each target molecule, the diffusion data of the molecule were collected from 3,500,000 different initial states.

Figure 1b shows the probability distribution of the methanol molecule which has a position vector $\vec{s}(t)$ with a position angle $\theta$, denoted by $P(\theta, t)$, for different time $t$. The value of $\theta$ is the angle between $\vec{s}(t)$ and the $z$ axis, as shown in Fig. 1a. The $z$ axis is defined as the origin orientation $\vec{i}(t = 0)$ of the methanol molecule, which is the direction pointing from the oxygen atom to the carbon atom. Remarkably, $P(\theta, t)$ shows a considerable orientation dependence. A peak at $\theta = 0$ can be clearly seen, indicating that the molecule prefers movement along the initial molecular orientation $\vec{i}(0)$. The difference $P_{diff}(t) = P(0, t) - P(\pi, t)$ increases from $t = 3$ ps, reaches a maximal of 16% at $t = 10$ ps, and then decreases. We note that the free diffusion of particle has been considered as completely random with isotropic probabilities along all directions [24-28] according to classical theory. The orientation-dependent diffusion we observed here clearly differs from this conventional viewpoint.

The distribution of the average displacement $S(\theta, t)$ with respect to $\theta$ is computed by $S(\theta,t) = \sum_\theta |\vec{s}(t)|/n_\theta$, where the sum runs over the samples that the position angles of $\vec{s}(t)$ fall in the interval of [$\theta$-$\Delta\theta/2$, $\theta$+$\Delta\theta/2$], and $n_\theta$ is the number of such samples. We set $\Delta\theta = \pi/18$. $S(\theta, t)$ has a weak orientation-dependence (see Fig. 1c). In order to include both the effects of the average displacement and probability, we calculate the normalized accumulated displacements, defined as $\Theta(\theta,t) = P(\theta,t) \cdot S(\theta,t)$. $\Theta(\theta,t)$ represents the accumulated displacements of the methanol molecule in all samples with position angle of $\theta$ (the direct definition of $\Theta(\theta,t)$ from $\vec{s}(t)$ can be seen in Appendix), and normalized to make $\overline{\Theta}(t)$ equal to the average displacement $\langle|\vec{s}(t)|\rangle$ (here, $\langle\cdots\rangle$ is the average for all the samples, and $\overline{\Theta}(t) = \int_0^\pi \Theta(\theta,t)d\theta/\pi$). Both high probability and a large average displacement at angle $\theta$ result in a larger $\Theta(\theta,t)$. $\Theta(\theta,t)$ has a unit of *length•probability*. In order to emphasize the contribution of the orientation-dependent part in the diffusion and considering that $P_{diff}(t)$ is ~8% (see inset in Fig. 1b), we use nm•8% as the unit of $\Theta(\theta,t)$. Fig. 1d shows that the $\Theta(\theta,t)$ curves deviate considerably from circles, further indicating the preponderance of the molecule to move along the initial orientation.

To more quantitatively characterize the asymmetric behavior of diffusion, we measure the difference of the maximal and minimal values of $\Theta(\theta, t)$ as

$$\Theta_{\text{diff}}(t) = \Theta(0, t) - \Theta(\pi, t) \qquad . \quad (1)$$

From Fig. 1e, we can see that $\Theta_{\text{diff}}(t)$ increases gradually from $t = 0$, and reaches a plateau value (1.0 nm•8%) after 30 ps with some fluctuations.

This plateau results from the different probability distribution and average displacements at different times. For example, at $t = 20$ ps, there are 13% more probabilities for the molecule to move along the initial molecular orientation than opposite to the orientation with a small average displacements (both $S(0, t)$ and $S(\pi, t)$) ~0.5 nm. At $t = 100$ ps, only 8% more probabilities for the molecule to move along the initial molecular orientation and the average displacement reaches ~1.2 nm (see Fig. 1c). We note that it is the use of nm•8% as the unit of $\Theta(\theta, t)$ so that the orientation-dependent part in diffusion can be appropriately characterized by $\Theta_{\text{diff}}(t)$. The simple use of 'nm' is possible to misinterpret the importance of the orientation-dependent part, appearing as a small value of ~1 Å, which is covered up because of the average with the orientation-free part.

Next, we explore the physics behind this observation. We postulate that the asymmetry of the molecule is the key to this unexpected observation. In order to reveal the effect of the molecular asymmetry, we pulled a methanol molecule along and opposite to the z axis with a constant force of $F$ = 5.32 pN and collected the displacements as shown in Fig. 1f. The linear fits of the data were applied to obtain the damping coefficient of 1.88×10$^{-12}$ kg/s and 1.99×10$^{-12}$ kg/s, respectively, for moving forward and backward. It is clear that the damping coefficient of moving forward is considerably less than the coefficient of moving backward. Assuming that we can confine the molecule to move freely only along the orientation (z-direction), we expect that the molecule will have a net movement along z direction under thermal fluctuation. Thus, in a finite time period, the molecule will have a net forward displacement $\Theta_{\text{diff}}(t) > 0$, since it requires time to regulate the orientation of the molecule from its initial orientation (+z). It is readily apparent that the auto-correlation of molecule orientation angle $\varphi$, denoted as $C_\varphi(t) = \langle \vec{i}(t) \cdot \vec{i}(0) \rangle = \langle \cos \varphi(t) \rangle$, is crucial for the accumulation of $\Theta_{\text{diff}}(\theta, t)$ in the time period of $t$. A positive value of $C_\varphi(t)$ indicates the upward preponderance of the molecule. In Fig.1e, it can be clearly seen that, during the first 30 ps, $C_\varphi(t)$ is positive and gradually decays to zero, while $\Theta_{\text{diff}}(t)$ increases and finally reaches a plateau. We note that the asymmetric part of diffusion we observed is considerably large, and the proportion $\Theta_{\text{diff}}(t)/\langle |\vec{s}(t)| \rangle$ reaches a maximum at 18% at $t \sim$ 10 ps (at the same time, $P_{diff}(t)$ reaches 16%, see Fig. 1b and 1g). As $t$ increases, $\langle |\vec{s}(t)| \rangle$ increases monotonically while $\Theta_{\text{diff}}(t)$ is saturated. This proportion becomes smaller and eventually negligible in macroscopic timescale.

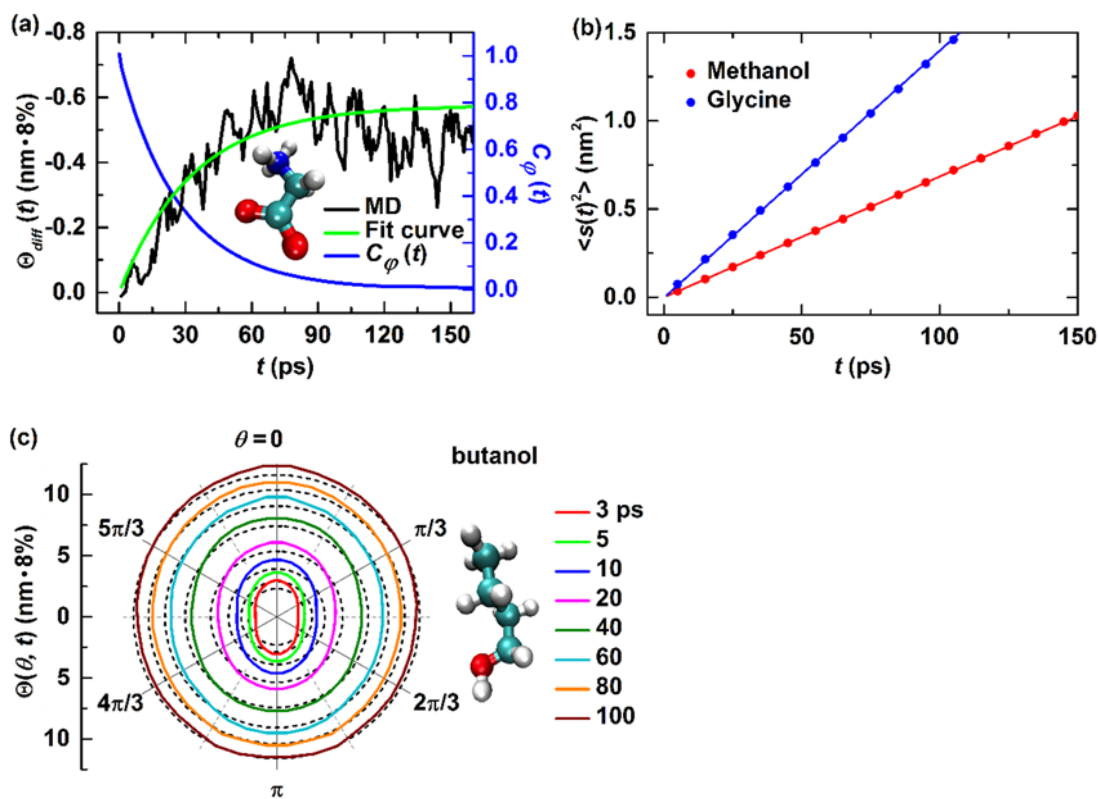

**Figure 2.** Orientation-dependent diffusion and characteristic time for the increase of $\Theta_{diff}(t)$, and the decay of $C_\varphi(t)$ in finite timescales. **(a)** Difference of the forward and backward displacement $\Theta_{diff}(t)$, together with the average orientation angle (green) at time $t$ for (a) the glycine in water solutions. The color setting of curves is the same as in the above Fig. 1f. **(b)** Mean square displacement with respect to $t$, showing the relation predicted classically $\langle \vec{s}(t)^2 \rangle = 6Dt$ still stands. The diffusion constants $D$ for the methnal, glycine and model molecules are 0.1 Å²/ps, 0.2 Å²/ps, and 0.8 Å²/ps, respectively. **(c)** Normalized accumulated displacement $\Theta(\theta, t)$ of butanol molecule. The left vertical axis is the radius coordinate. Different colors represent different t time. The right part is obtained by mirror imaging. The circles correspond to $\overline{\Theta}(t)$, for guiding the eyes. The initial orientations of the butanol

molecule is defined as the directions pointing from the oxygen atom to the methyl group.

The existence of the orientation-dependent diffusion within finite time periods is universal. We have also performed simulations for other molecules including glycine in water solutions and a model molecule solvated in LJ particles. As shown in Fig. 2a, the increase of $\Theta_{\text{diff}}(t)$ is clearly correlated with the decrease of $C_\varphi(t)$, whereas $\Theta_{\text{diff}}(t)$ increase exponentially and $C_\varphi(t)$ decreases exponentially. We note that, for some other molecules, that is ethanol, propanol, and butanol, the most significant difference on the diffusion may not be along the chain. For example, the diffusion along the chain is much larger than the diffusion perpendicular the chain for butanol (see Fig. 2c). However, the asymmetrical displacement can still be seen clearly.

Finally, we checked whether the diffusion still satisfied $\langle \vec{s}(t)^2 \rangle = 6Dt$ as predicted by Einstein [1], if we did not consider the asymmetry in the diffusion. Figure 2b shows the good linear dependence of $\langle \vec{s}(t)^2 \rangle$ on time *t*.

In summary, in finite timescales of picoseconds to nanoseconds, the diffusion of an asymmetric nanoparticle (including molecule) includes two parts:

*Free diffusion = Isotropic diffusion ⊕ Orientation-dependent diffusion.*

The latter part can reach more than 10% of the former part, and results from the orientation-dependent damping force of the molecule together with a finite time required to regulate the molecular orientation. This finding extends the work of Einstein to nano-world beyond random Brownian motion, and is expected to have an essential role in the understanding of the nanoscale world. We also expect that directional transportation in macroscopic to nanoscopic time-scale such as biased water transports through nanochannels [29-32] can be accumulated from the orientation-dependent displacements of nanoparticles by controlling the particle orientation with various methods such as electric and magnetic fields. This provides new ideas to develop novel technology for various nanoscale and bulk applications, in such things as chemical separation, sensing and drug delivery.

Acknowledgments

We thank Qing Ji and Jun Hu for helpful discussions. This work was partially supported by the National Natural Science Foundation of China under Grant Nos. 10825520, 11105088, 11175230 and 11290164. We gratefully acknowledge the support from Shanghai Supercomputer Center and Supercomputing Center of Chinese Academy of Sciences.

**Appendix**

**Definition of the normalized accumulated displacement $\Theta(\theta, t)$ directly to $\vec{s}(t)$ and its relationship with $S(\theta, t)$ and $P(\theta, t)$**

$\Theta(\theta, t) = C \sum_\theta |\vec{s}(t)|/n\Omega_{\theta,\Delta\theta}$ with the sum running over all the samples that the orientation angles of $\vec{s}(t)$ falling in the interval of [$\theta$-$\Delta\theta$/2, $\theta$+$\Delta\theta$/2]. The value of $n$ is the number of total samples while $\Omega_{\theta,\Delta\theta}$ is a normalization factor for the solid angle, $\Omega_{\theta,\Delta\theta} = \sin\theta\sin\Delta\theta/2$. $C$ is the normalized factor to make $\overline{\Theta}(t)$ equal to the average displacement of the molecule for all samples $\langle|\vec{s}(t)|\rangle$ (here $\langle\cdots\rangle$ is the average for all the samples, and $\overline{\Theta}(t) = \int_0^\pi \Theta(\theta,t)d\theta/\pi$). The value of $\Theta(\theta,t)$ represents the accumulated displacements of the molecule for all the samples with a position angles $\theta$ since the denominator is independent to the number of the samples with the position angles $\theta$. The existence of $\Omega_{\theta,\Delta\theta}$ in the denominator results from the projection of the 3D space to the 2D place spanned by z axis and orientation angle $\theta$.

$$\Theta(\theta,t) = C\sum_\theta |\vec{s}(t)|/n\Omega_{\theta,\Delta\theta} = \sum_\theta |\vec{s}(t)|/n_\theta \cdot Cn_\theta/n\Omega_{\theta,\Delta\theta} = S(\theta,t) \cdot P(\theta,t)$$

since $S(\theta,t) = \sum_\theta |\vec{s}(t)|/n_\theta$, where the sum runs over all the samples such that the orientation angles of $\vec{s}(t)$ fall in the interval of $[\theta-\Delta\theta/2, \theta+\Delta\theta/2]$, and $n_\theta$ is the number of such cases. Thus, $P(\theta,t) = Cn_\theta/n\Omega_{\theta,\Delta\theta}$ with the normalized factor $C$ determined by $\int_0^\pi P(\theta,t)d\theta/\pi = 1$.